\documentstyle[12pt,a4wide,epsf]{article}
\newcommand{\nin}{\noindent}
\newcommand{\be}{\begin{equation}}
\newcommand{\ee}{\end{equation}}
\newcommand{\bea}{\begin{eqnarray}}
\newcommand{\eea}{\end{eqnarray}}

\newcommand{\hf}{\frac{1}{2}}
\newcommand{\nn}{\nonumber\\}

\begin{document}

%\nin KCL-PH-TH/2013-{\bf 09}

\begin{center}
{\bf{\large Galaxy Rotation Curves in Covariant Ho\v{r}ava-Lifshitz Gravity}

\vspace{0.5cm}

J. Alexandre\footnote{jean.alexandre@kcl.ac.uk}, M. Kostacinska\footnote{martyna.kostacinska13@imperial.ac.uk}}\\
$^1$ King's College London, Department of Physics, WC2R 2LS\\
$^2$ Imperial College London, Theoretical Physics, SW7 2AZ

\vspace{2cm}

{\bf Abstract}

\end{center}

\vspace{0.5cm}

\nin Using the multiplicity of solutions for the projectable case of the covariant extension of Ho\v{r}ava-Lifshitz Gravity, 
we show that an appropriate choice for the auxiliary field allows for an effective description of galaxy rotation curves. 
This description is based on static and spherically symmetric solutions of covariant Ho\v{r}ava-Lifshitz Gravity and does not 
require Dark Matter.

\section{Introduction}

Although General Relativity (GR) has been accurately checked in the solar system, the small distance behaviour of gravity is still not 
understood, and several alternatives to GR have been proposed. Amongst these 
modified theories is Ho\v{r}ava-Lifshitz (HL) gravity \cite{Horava}, which is based on the Lifshitz approach and consists of 
introducing an anisotropy between space and time: when space is rescaled as $x\to bx$, time is rescaled as 
$t\to b^zt$, where $z$ is an integer. In Lifshitz models, Lorentz invariance must be recovered for $z=1$, and the situation $z\ne1$
leads to new renormalizable interactions: the convergence of loop integrals is improved by inclusion of higher orders in space derivatives, 
without introducing ghosts, since the number of time derivatives remains the minimum. 
A review of Lifshitz-type quantum field theories in flat space time can be found in \cite{reviewflat}, 
and the HL power-counting renormalizable theory of gravity is reviewed in \cite{reviewHL1}, \cite{reviewHL2} and \cite{reviewHL3}.

A fundamental problem of the original model of HL gravity is the existence of an additional scalar degree of freedom for the metric, which
can be understood as a Goldstone mode arising from breaking of 4-dimensional diffeomorphism \cite{strongcoupling1}, \cite{strongcoupling2}.
A solution to this problem has been proposed in \cite{covariantHL}, where an auxiliary field $A$ is introduced, such that its ``equation of motion''
leads to an additional constraint, eliminating the unwanted scalar degree of freedom of the metric. 
The resulting theory is invariant under a new Abelian gauge symmetry $U_\Sigma(1)$ 
which involves the metric components, the auxiliary field $A$ and an additional auxiliary field $\nu$. This gauge symmetry, together with 
the 3-dimensional diffeomorphism of HL gravity, can be shown to be equivalent to a 4-dimensional diffeomorphism at the lowest order in a 
post-Newtonian expansion, showing the equivalence with GR at long distances. 
For this reason, this modified version of HL gravity is called covariant HL gravity. 
We note however, that the long-distance limit is not obviously recovered: it has been shown in \cite{dasilva2} that the equivalence principle
is not automatically retrieved in the infrared, and that the meaning of the auxiliary
field $A$ and its coupling to matter are still open questions.

Because of the anisotropy between space and time, HL gravity is naturally described in terms of the Arnowitt-Deser-Misner (ADM) 
decomposition of the metric, which expresses a space-time foliation. An important consequences of space-time anisotropy is the possibility
of imposing the lapse function $N$ to depend on time only. This situation is called the projectable case
and leads to an interesting feature: the solutions of the equations of motion are not unique. Indeed, the Hamiltonian constraint, obtained
by variation of the action with respect to $N$, leads to an integral equation, which does not have a unique 
solution, as will be seen in the present article.
This multiplicity of solutions is independent of the above mentioned new gauge symmetry $U_\Sigma(1)$, and 
the different solutions for $A$ in the projectable case belong to different gauge orbits.
On the other hand, in the non-projectable case where $N$ depends on both space and time, the Hamiltonian constraint leads to a differential
equation, which has a unique solution, after fixing the constants of integration. 
The non-projectable case has been studied in \cite{non-projectable} for static spherically symmetric solutions of covariant HL gravity.

\vspace{0.5cm}
 
In this article we use the freedom of choice of the auxiliary field $A$ in the projectable case 
to study the possibility of fitting galaxy rotations curves without the need for Dark Matter. One can argue that the ambiguity in the 
choice of $A$ leaves its physical interpretation unclear. This may be understood as one of the incomplete aspects of covariant HL gravity, and
one could conjecture the existence of an additional symmetry which would 
restrict the set of solutions for the auxiliary field $A$, but this still remains to be studied. 
We note that the multiplicity of solutions, arising from a Hamiltonian constraint expressed in terms of an integral over space, has been 
discussed in \cite{mukohyama} as an alternative to Dark Matter, in the context of the original HL gravity.
Fits to galaxy rotation curves have been studied in the framework of the original HL gravity in \cite{HLrotation1} and \cite{HLrotation2}, where a detailed
comparison with experimental data is presented.
Also, an interesting analogy between HL gravity for $z=0$ and Modified Newtonian Dynamics theories has been described in \cite{HLMOND}.

\vspace{0.5cm}

In section 2 we review the static and spherically symmetric solutions of covariant HL gravity, and in section 3 we show how  
galaxy rotation curves can be described by the covariant extension of HL gravity. Although the curves we use were derived from Dark Matter models, 
we use them as experimental data fits. 
It is interesting to note that we obtain exact solutions for the auxiliary field $A$. Another exact solution in the Lifshitz context has been
derived in \cite{Liouville}, where the exact effective potential for a Liouville scalar theory, renormalizable in 3+1 dimensions with 
anisotropic scaling $z=3$, proves to be an exponential, as in the usual 1+1 dimensional relativistic Liouville theory.
Finally, we conclude by discussing the possiblity to describe
the solar system with our solution.

\section{Static and spherically symmetric solutions of covariant HL gravity}

The static and spherically symmetric solutions of covariant HL gravity were derived in \cite{solutions1} and \cite{solutions2}, and we review here the aspects 
relevant to our present study.

\subsection{Action}

The ADM metric we consider is
\be
ds^2=-  c^2 N^2 dt^2+g_{ij} \left(dx^i+N^i dt\right)\left(dx^j+N^jdt\right)~,
\ee
where $N$ and $N^i$ are the lapse and shift functions respectively, and $g_{ij}$ is the three-dimensional space metric.   
We consider the anisotropic scaling with $z=3$, for which the different operators, allowed for the model to be power-counting renormalizable,
have the mass dimension of 6, at most. We note that in this case the mass dimension of the speed of light is 
$[c]=z-1=2$, and we will keep $c$ in the different expression for the sake of clarity.

\nin The covariant Ho\v{r}ava-Lifshitz action, in the absence of cosmological constant, is
\bea\label{S}
S&=&\int dt d^3x\sqrt{g}\left( N\left[ K_{ij}K^{ij}-K^2-V+\nu\Theta^{ij}(2K_{ij}+\nabla_i\nabla_j\nu)\right]-AR^{(3)}\right)~,
\eea
where $g$ is the determinant of the three-dimensional metric $g_{ij}$, with curvature scalar $R^{(3)}$.
$A$ is an auxiliary field (without kinetic term) of mass dimension $z+1=4$ and $\nu$ is an auxiliary field of mass dimension $z-2=1$. 
In the above expression, the extrinsic curvature is
\be\label{Kij}
K_{ij}=\frac{1}{2 N} \left\{\dot{g}_{ij}-\nabla_i N_j-\nabla_j N_i\right\},~~ i,j=1,2,3~,
\ee
 where a dot denotes a time derivative, and
\be
\Theta^{ij}=R^{(3)ij}-\frac{1}{2}R^{(3)} g^{ij}~,
\ee
where where $R^{(3)ij}$ is the Ricci tensor corresponding to the three-dimensional space metric $g_{ij}$.
Note that a more general kinetic term $K_{ij}K^{ij}-\lambda K^2$ can be considered, with $\lambda$ a generic parameter. 
The latter should be equal to 1 if one wishes to recover GR in the infrared, but in the framework of HL gravity, 
no symmetry imposes this parameter to be equal to 1. Constraints on $\lambda$ resulting from observations of Type 
Ia Supernovae, Baryon Acoustic Oscillations, CMB and the requirement of Big Bang Nucleosynthesis all point towards 
a value near the GR parameter $\lambda = 1$ \cite{Dutta:2010jh}, while it has also been shown that covariant HL gravity with $\lambda\ne1$ 
leads to inconsistencies, if one compares the predictions of the model with solar system tests \cite{solarsystem}.
We therefore focus here on the situation where $\lambda=1$ only. Independently of these tests, the mathematical consistency of covariant HL 
gravity for $\lambda\ne1$ has been studied in \cite{dasilva1}, and the corresponding Hamiltonian structure in \cite{kluson}.
We stress again here that the purpose of this covariant extension of HL gravity is to eliminate the unwanted additional scalar graviton
from the spectrum. As shown in \cite{covariantHL}, no singularity arises in the limit $\lambda\to1$, and the 
strong coupling of the scalar graviton does not exist anymore, as it is the case for the original HL gravity.
Besides \cite{strongcoupling1} and \cite{strongcoupling2}, one can find detailed analysis of the strong coupling problem 
in \cite{lambda=11}, \cite{lambda=12}, \cite{lambda=13},
and also related discussions in \cite{discussions} on a extension of the original HL gravity \cite{extension}.

The potential term $V$ contains up to six spatial derivatives of the metric $g_{ij}$:
\bea 
V&=&-c^2 R^{(3)}-\alpha_1 (R^{(3)})^2-\alpha_2 R^{(3)ij}R_{ij}^{(3)}-\beta_1 (R^{(3)})^3-\beta_2 R^{(3)} R^{(3)ij} R_{ij}^{(3)}\\
&&-\beta_3 R_j^{(3)i} R_k^{(3)j} R_i^{(3)k}
-\beta_4 R^{(3)}\nabla^{2}R^{(3)}-\beta_5 \nabla^i R_{jk}^{(3)} \nabla_i R^{(3)jk}~,
\eea
where the mass dimensions of the different couplings are 
\be
[\alpha_1]=[\alpha_2]=z-1=2~~~,~~~[\beta_1]=[\beta_2]=[\beta_3]=[\beta_4]=[\beta_5]=z-3=0~.
\ee
We consider the most general, static, spherically symmetric metric, of the form:
\be\label{metric}
ds^2=-c^2 N^2 dt^2+\frac{1}{f(r)}\left(dr+n(r)dt\right)^2+r^2( d\theta^2+\sin^2\theta d\phi^2),
\ee
where $n(r)=N^{r}(r)$ is the radial component of shift function, and $N_r=n(r)/f(r)$ since $g_{rr}=1/f(r)$. Note that $n$
has mass dimension $z-1=2$.

\subsection{Constraints and equations of motion}

The variation of the action (\ref{S}) with respect to the different degrees of freedom leads to the following 
constraints, or equations of motion, where a prime denotes a derivative with respect to the radial coordinate $r$.

\begin{itemize}

\item The variation with respect to $A$ gives $R^{(3)}=0$, or equivalently
\be
r f'+f-1=0~,
\ee
which imposes the function $f(r)$ in the metric (\ref{metric}) to be
\be\label{solf}
f(r)=1-\frac{2 B}{r}~,
\ee
where $B$ is a constant of integration.

\item The variation with respect to $\nu$ gives
\be \label{const2}
\Theta^{ij}\nabla_i\nabla_j\nu+\Theta^{ij}K_{ij}=0,
\ee
In what follows we will assume the Gauge fixing of $\nu=0$, then the above constraint gives $\Theta^{ij}K_{ij}=0$, which is satisfied for
spherically symmetric solutions.

\item The variation with respect to $N$ gives the Hamiltonian constraint
\be\label{Hamilton}
\int_{0}^{\infty}dr \frac{r^2}{\sqrt{f(r)}}(K_{ij}K^{ij}-K^2+V)=0~,
\ee
which is an space-integral equation since $N$ depends on time only. 

\item The variation with respect to $n$ gives
\be
f'(r) n(r)=0~,
\ee
such that either $f$ is a constant ($B=0$) or $n=0$.

\item The variation with respect to $f$ gives
\be\label{evoleqAr}
A'+\frac{A}{2r}\left(1-\frac{1}{f}\right)+4\frac{f n\left(\sqrt{r}n\right)'}{\sqrt{r}} = 
\left\{\frac{r}{4f}-\frac{\sqrt f}{2r}\sum_{n=0}^3(-1)^n
\frac{d^n}{dr^n}\left(\frac{r^2}{\sqrt f}\frac{\partial}{\partial f^{(n)}}\right)\right\}V,
\ee
where the time has been rescaled such that $N=1$.

\end{itemize}

\subsection{Solutions}\label{sol}

The different solutions of the previous set of equations were derived in \cite{solutions1}, \cite{solutions2}, and here we shortly review the different cases.
\begin{itemize}

\item \underline{$n=0$}: the solution of the differential equation (\ref{evoleqAr}) gives short-distance corrections to the Schwarzschild 
solution, details of which can be found in \cite{solutions1}, \cite{solutions2};

\item\underline{$n\ne0$ and $A=0$}: leads to the Schwarzschild solution in the Painleve-Gullstrand coordinates ($n\propto r^{-1/2}$), and
the Hamiltonian constraint is automatically satisfied;

\item\underline{$n\ne0$ and $A\ne0$}: corresponds to the situation with multiple solutions for $A$ and $n$, on which this article focuses.
Since $B=0$, we have $f=1$ and $V=0$, such that the equations (\ref{Hamilton}) and (\ref{evoleqAr}) give respectively 
\bea
\int_{0}^{\infty} dr(rn^2)'&=&0\nn
r A'+2(rn^2)'&=&0~,
\eea
with solutions
\be\label{nsquared}
n^2(r)=\frac{C}{r}-\frac{1}{2} A(r)+\frac{1}{2 r}\int_0^r d\rho~A(\rho)~,
\ee
where $C$ is a constant of integration and the auxiliary field $A$ satisfies
\be\label{intA}
\int_0^\infty dr  rA'(r)=0~.
\ee
The flexibility in the choice of solution for $A$ will help describe galaxy rotation curves, as explained in the next section.

\end{itemize}

\section{Galaxy rotation curves}

In what follows, we consider each star in the spiral arms of the galaxy as a test particle, moving on a circular trajectory under the influence of the 
potential $\phi(r)$, generated by the centre of the galaxy, which is assumed to be static and spherically symmetric. 
Our approach is to start from a stellar velocity distribution and derive the corresponding
expression for the auxiliary field $A$, in the above situation with $n\ne0$ and $A\ne0$ , where multiple solutions for $A$ are allowed. 
We note the following few points:

\nin{\it(i)} We consider vacuum solutions of covariant HL gravity, for which one cannot in principle describe 
the region in the centre of the galaxy. Nevertheless, as we will see in case of the following velocity profiles, a consistent solution for 
$A$ can be found for $0\leq r\leq R$, where $R$ is a typical radial length describing the galaxy. To be consistent with the usual studies of galaxy rotations, 
we will choose here $R=r_{200}$, corresponding to the virial radius of the galaxy. The latter is defined as the distance from the centre 
of the galaxy, where the density $\rho_{200}$ is 200 times the critical density $\rho_c$ of the Universe
\be
\rho_{200}=200\rho_c=200 \frac{3H^2}{8\pi G}~,
\ee
where $H$ is the Hubble constant;

\nin{\it(ii)} For $r>r_{200}$, we assume the solution $A=0$, which leads to the usual Schwarzschild solution outside the galaxy. 
Indeed, far from the galaxy,
one expects to see the Newtonian potential. This solution implies that the Hamiltonian constraint is expressed in terms of an
integral over the finite range $[0,r_{200}]$ of radial coordinate $r$, which we can impose to vanish by fixing the constants of integration;

\nin{\it(iii)} For $r\leq r_{200}$, since the shift $A\to A+$constant does not have a physical implication, 
one can expect to find a consistent solution for $A$ which is continuous at $r=r_{200}$, i.e. $A(r_{200})=0$. 
On the other hand, one cannot expect the auxiliary field to be
differentiable at $r=r_{200}$, but this is not necessary for the consistency of covariant HL gravity:
the second derivative of $A$ does not appear in any equation of motion or constraint, so the first derivative can be discontinuous.

\subsection{Gravitational potential}

On a circular trajectory, the relation between the speed $v$ of a star and the Newtonian potential $\phi$ is 
\be\label{v2/r}
\frac{v^2}{r}=\frac{d\phi}{dr}~,
\ee
where $\phi$ is obtained from the $g_{00}$ component of the low energy effective theory 
\be
-c^2+n^2=-c^2(1+2\phi)~,
\ee
and is thus given by
\be\label{phi}
\phi(r)=-\frac{n^2(r)}{2c^2}~.
\ee
We note here that, in the situation where the second auxiliary field $\nu$ does not vanish, the Newtonian potential is instead given by
\be
\phi(r)=-\frac{\left(n(r)-\nabla_r \nu(r)\right)^2}{2 c^2}~,
\ee  
and $\phi$ is independent of the $U_\Sigma(1)$ gauge choice for $\nu$ and $A$. Also, $\phi$ is dimensionless, so that the speed $v$ we use in this 
article is also dimensionless, which corresponds to a usual definition of speed, for isotropic space-time. The corresponding ``Lifshitz velocity''
is $cv$, with mass dimension $z-1=2$.\\
From the relations (\ref{nsquared}) and (\ref{v2/r}), the speed $v$ and the auxiliary field $A$ are then related by
\be\label{rA'}
\frac{rA'(r)}{4c^2}=\frac{d}{dr}\left(r\int^r d\rho\frac{v^2(\rho)}{\rho}\right)=v^2(r)+\int^r d\rho\frac{v^2(\rho)}{\rho}~,
\ee
and we have to check that the Hamiltonian constraint (\ref{intA}) is satisfied, in order to show the consistency of the approach.\\
We also note from eqs.(\ref{nsquared},\ref{v2/r}) that the solution $A=0$ gives, 
\be
v(r)=\sqrt\frac{C}{2c^2r}~,
\ee
which is expected from Newtonian mechanics outside the galaxy, for $r>r_{200}$. If one fixes the constant of integration $C$ to the value
\be
C=2c^2r_{200}v^2_{200}~,
\ee
where $v_{200}=v(r_{200})$, one obtains a continuous speed at $r=r_{200}$. 
Finally a different choice of the auxiliary field $\nu$ would modify the field $A$ in such a 
way that physical results would not be changed.

\subsection{Navarro, Frenk and White profile}

This profile originates from Cold Dark Matter halo models and is characterised by the ``cusp" shape of the density distribution. 
Its validity has been tested on variety of observational results, including Low Surface Brightness galaxies in \cite{LSB} and spherical 
galaxies and clusters in \cite{Lokas}. For our purposes, we use the circular velocity profile derived from the mass density profile in 
\cite{NFW}, based on the assumption of a Dark Matter halo and ignoring the contribution of baryons, 
\be\label{NFWprofile}
v(r)=v_0\sqrt{\frac{r_{200}}{r}\ln\left(1+\frac{ar}{r_{200}}\right)-\frac{a}{1+ar/r_{200}}}~,
\ee
where $a$ is the concentration parameter and
\be
v_0=\frac{v_{200}}{\sqrt{\ln{(1+a)}-a/(1+a)}}
\ee
where $v_{200}$, the virial velocity, is the circular velocity at the virial radius. Some typical values are \cite{deBlok}

\begin{center}
\begin{tabular}{p{1.9cm}|p{1.9cm}p{2.5cm}p{1.9cm}}
	Galaxy & $a$ &$v_{200}$ [km/s]&$r_{200}$ [kpc] \\ \hline
	NGC 2403 &$10.9 \pm 0.6$ &$106.1 \pm 1.9$ &$88.4 \pm 2.6$ \\ 
	NGC 3198 &$11.2\pm 0.43$ &$104.0 \pm 0.7$ &$86.6 \pm1.6$ \\ 
	NGC 3521 &$14.0\pm 12.6$ &$122.5 \pm 20.4$ &$102.0 \pm 18$ \\ 
\end{tabular}
\end{center}

\nin With the profile (\ref{NFWprofile}), eq.(\ref{rA'}) leads to
\be
\frac{rA'(r)}{4c^2v_0^2}=\phi_1-\frac{a}{1+ar/r_{200}}~,
\ee
where $\phi_1$ is a constant of integration, and therefore
\be
\frac{A(r)}{4c^2v_0^2}=(\phi_1-a)\ln\left(\frac{r}{r_{200}}\right)+a\ln\left(1+a\frac{r}{r_{200}}\right)+A_1~,
\ee
where $A_1$ is a constant of integration. Since we consider the solution $A(r)=0$ for $r>r_{200}$ and impose $A(r)$ to be
continuous at $r=r_{200}$, then $A_1=-a\ln(1+a)$ and 
\bea\label{ANFW}
\frac{A(r)}{4c^2v_0^2}&=&(\phi_1-a)\ln\left(\frac{r}{r_{200}}\right)+a\ln\left(\frac{1+ar/r_{200}}{1+a}\right)~~~\mbox{if}~~~0\leq r\leq r_{200}\\
A(r)&=&0~~~\mbox{if}~~~r>r_{200}~.\nonumber
\eea
Finally, the Hamiltonian constraint
\be
0=\int_0^\infty dr~rA'(r)=4c^2v_0^2\int_0^{r_{200}} dr\left(\phi_1-\frac{a}{1+ar/r_{200}}\right)~,
\ee 
is satisfied if
\be
\phi_1=\ln(1+a)~.
\ee
We note that the solution (\ref{ANFW}) is consistent for any relevant values of the parameters $a,v_0,r_{200}$ 
and therefore allows the description of a whole range of galaxies.

\subsection{Pseudoisothermal profile} 

The pseudoisothermal mass density profile assumes an existence of a ``cored" dark matter halo component in the galaxy \cite{LyndenBell} , 
with mass density of an approximately constant value in the central region of the galaxy, for $r\leq R_c$. 
Empirically motivated, the model is often contrasted with the aforementioned NFW ``cuspy" profile, while evidence shows that 
it provides better fit of the galaxy rotation velocities \cite{LSB}. The velocity profile is  
\be\label{pseudoprofile}
v(r)=v_0\sqrt{1-\frac{R_C}{r}\arctan\left(\frac{r}{R_C}\right)}~,
\ee
where $v_0$ is given by
\be
v_0=\sqrt{4\pi G \rho_0 R_C^2}~,
\ee
and $\rho_0$, the central density, is the density within $R_C$. Typical values are \cite{deBlok}

\begin{center}
\begin{tabular}{p{2cm}|p{2cm}p{3cm}p{2.5cm}} 
	Galaxy  &$R_c$ [kpc] &$\rho_0$ [$10^{-3}$M$_\odot$pc$^{-3}$] & $r_{200}$ [kpc]\\ \hline
	NGC 2403 &$2.51\pm 0.32$  &$59.1 \pm 14.3$ & $88.4 \pm 2.6$\\ 
	NGC 2841 &$1.36\pm 0.75$  &$674.8 \pm 736.4$ & $159.0 \pm 3.7$\\ 
	NGC 3621 &$5.88\pm 0.32$  &$13.0 \pm 1.1$ &$106.6 \pm 4.0$\\
\end{tabular}
\end{center}

\nin We note that in the case of NCG 2403, both the NFW and the present profile can be shown to produce fits of comparably good quality. \\
With the profile (\ref{pseudoprofile}), eq.(\ref{rA'}) leads to
\be
\frac{rA'(r)}{4c^2v_0^2}=\phi_2+\frac{1}{2}\ln\left(1+\frac{r^2}{R_c^2}\right)~,
\ee
where $\phi_2$ is a constant of integration. 
The same steps as those describe for the previous profile lead to the solution
\bea\label{Apseudo}
\frac{A(r)}{4c^2v_0^2}&=&\phi_2\ln\left(\frac{r}{r_{200}}\right)+\hf\int_0^{r/R_c}\frac{dt}{t}\ln(1+t^2)
+A_2~~~\mbox{if}~~~0\leq r\leq r_{200}\\
A(r)&=&0~~~\mbox{if}~~~r>r_{200}~,\nonumber
\eea
where the constants of integration are 
\be
A_2=-\hf\int_0^{r_{200}/R_c}\frac{dt}{t}\ln(1+t^2)~,
\ee
for the continuity of $A(r)$ at $r=r_{200}$, and
\be
\phi_2=-\hf\ln\left(1+\frac{r_{200}^2}{R_c^2}\right)+1-\frac{R_c}{r_{200}}\arctan\left(\frac{r_{200}}{R_c}\right)~,
\ee
for the Hamiltonian constraint to be satisfied.

\subsection{Vacuum energy contribution of the auxiliary field}

Since the present model does not require Dark Matter to explain galaxy rotation curves, the resulting mass content of the Universe might therefore be 
reduced. But the static configuration of $A$ actually results in an effective vacuum energy. This can be seen from the original action (\ref{S}), 
after integrating by parts the term $-\sqrt{g}AR^{(3)}$, which involve second-order space derivatives of the metric. 
This leads to the following contribution to the Hamiltonian  
\be\label{mass}
M\propto\frac{1}{c}\int d^3x\sqrt{g}g_{ij}\nabla^i\nabla^jA=\frac{4\pi}{c}\int_0^{r_{200}} r^2dr~\partial_r^2A~,
\ee
where the factor $1/c$ arises from the absence of integration over time, and restores the correct dimension.\\
In the case of the Navarro, Frenk and White profile, the energy (\ref{mass}) reads
\be
M_1=16\pi cv_0^2 r_{200}\left[\ln(1+a)-\frac{a}{1+a}\right]~,
\ee
which is always positive, for any value of the parameter $a\ge0$.\\
For the Pseudoisothermal case, we obtain from eq.(\ref{mass})
\be
M_2=16\pi cv_0^2 r_{200}\left[1-\frac{R_c}{r_{200}}\arctan\left(\frac{r_{200}}{R_c}\right)\right]~,
\ee
which is also always positive, for any value of the ratio $r_{200}/R_c$.\\
The auxilliary field $A$ has therefore a cosmological role, by contributing to the vacuum energy of the Universe, consistently with its observed acceleration.

\section{Conclusion}

The multiplicity of solutions in covariant HL gravity has been used to describe galaxy rotation curves, without the need for Dark Matter. 
This multiplicity is a consequence of the breaking of 4-dimensional diffeomorphism, in the projectable case, where the lapse function is 
imposed to depend on time only.
We considered two specific velocity profiles and showed in each case that a corresponding consistent solution can 
be derived for the auxiliary field $A$. Clearly, similar results can be expected with other profiles.
A more detailed analysis would consist in taking into account the finite star density at the centre of the galaxy, as well as its rotation.

One should note that the solutions found, for the auxilliary field $A$, depend on parameters wich cannot appear in the Lagrangian in a universal way. 
These parameters 
arise as constant of integrations, as a result of the integral form of the Hamiltonian constraint, and they have to be fitted to each galaxy which is studied.
This is a bit similar to the Schwazschild metric in General Relativity, involving a mass which is a constant of integration and 
depends on the star/black hole under consideration.

It is interesting to note that the solutions (\ref{ANFW}) and (\ref{Apseudo}) actually do not only describe galaxy rotations curves, 
but also the usual Schwarzschild solutions for the Solar System, by taking the limit $r_{200}\to0$ such that $A=0$ for all $r$.
In this sense these solutions provide an effective unified description of gravitational effects, from the solar system scale to the
galaxy scale, where local Lorentz violation appears gradually, as the observational length increases from the the size of the solar system
to the size of the galaxy.

We finally note that, although HL gravity was initially developed to modify ultraviolet behaviour of gravity, our present work uses another 
consequence of local Lorentz symmetry violation, which is the existence of a non-unique solution to the Hamiltonian constraint. 
Since it has been shown that canonical Ho\v{r}ava - Lifshitz gravity provides a description of Cosmology which is consistent with 
constraints from observational data \cite{Dutta:2009jn}, it would be interesting to see if the non-unique time-dependent solutions of 
the covariant version of the theory could also allow for an effective description of Cosmology, including an alternative to Dark Energy.

\end{document}